\documentclass[11pt,preprint]{aastex}
\newcommand{\be}{\begin{equation}}
\newcommand{\ba}{\begin{eqnarray}}
\newcommand{\ee}{\end{equation}}
\newcommand{\ea}{\end{eqnarray}}
\newcommand{\etal}{et al.\ }
\newcommand{\etalb}{et al.}

\newcommand{\Ni}{N_{\rm ion}}
\newcommand{\Omm}{\Omega_m}
\newcommand{\Oml}{{\Omega_{\Lambda}}}

\newcommand{\Lya}{\mbox{Ly$\alpha$} }
\newcommand{\Ly}{\mbox{Ly$\alpha$}}

\begin{document}
\title{Was the Universe Reionized at Redshift 10?}

\author{Abraham Loeb} 
\affil{Astronomy Department, Harvard University, 60 
Garden Street, Cambridge, MA 02138; aloeb@cfa.harvard.edu}

\author{Rennan Barkana}
\affil{School of Physics and Astronomy, The Raymond and Beverly Sackler
Faculty of Exact Sciences, \\
Tel Aviv University, Tel Aviv 69978, ISRAEL; barkana@wise.tau.ac.il}

\author{Lars Hernquist} 
\affil{Astronomy Department, Harvard University, 60 
Garden Street, Cambridge, MA 02138; lars@cfa.harvard.edu}

\begin{abstract}

Recently, \citet{z10} claimed to have discovered a galaxy at a redshift
$z=10$, and identified a feature in its spectrum with a hydrogen Ly$\alpha$
emission line. If this identification is correct, we show that the
intergalactic medium (IGM) around the galaxy must be significantly ionized;
otherwise, the damping wing of Ly$\alpha$ absorption by the neutral IGM
would have greatly suppressed the emission line.  We find either that the
large-scale region surrounding this galaxy must have been largely reionized
by $z=10$ (with a neutral fraction $\la 0.4$) or that the stars within the
galaxy must be massive ($\ga 100\, M_\odot$), and hence capable of
producing a sufficiently large \ion{H}{2} region around it. We generalize
these conclusions and derive the maximum \Lya line flux for a given UV
continuum flux of galaxies prior to the epoch of reionization.

\end{abstract}

Key Words: galaxies: high-redshift, cosmology: theory, 
galaxies: formation

\section{Introduction}

\citet{z10} have reported the discovery of a galaxy at a
redshift $z=10$, which is gravitationally magnified by a factor $\sim
25-100$ by the foreground galaxy cluster Abel 1835. The redshift
identification is supported by the spectroscopic detection of an
emission line at $1.338\mu$m, the photometric signature of an
absorption trough at shorter wavelengths, and the lensing geometry.
Here, we examine the far-reaching implications of associating this
spectral feature with the Ly$\alpha$ emission line from a galaxy at
$z=10$. In particular, we show that if hydrogen in the intergalactic
medium (IGM) around this galaxy were neutral, then the damping wing of
its \Lya absorption would have greatly suppressed the \Lya emission
line of the galaxy.  Our results provide a strong incentive for
obtaining higher quality data on this galaxy.  More generally, we
derive the maximum \Lya line flux for a given UV continuum flux of
galaxies prior to the epoch of reionization. Future detection of
bright \Lya lines from high-redshift galaxies can be used to improve
the current lower limit on the redshift of reionization.

Throughout, we adopt values for cosmological parameters derived from
recent observations of the cosmic microwave background \citep{WMAP,
uros}. For the contributions to the energy density, we assume ratios
relative to the critical density of $\Omm=0.27$, $\Oml=0.73$, and
$\Omega_b=0.046$, for matter, vacuum energy, and baryons,
respectively.  We also choose a Hubble constant $H_0=72\mbox{ km
s}^{-1}\mbox{Mpc}^{-1}$, and a primordial scale-invariant ($n=1$)
power-law power spectrum with $\sigma_8=0.9$, where $\sigma_8$ is the
root-mean-square amplitude of mass fluctuations in spheres of radius
$8\ h^{-1}$ Mpc.

\section{Basic Absorption Parameters}

The absorption profile owing to \ion{H}{1} in the IGM depends on the
redshift $z_S$ and halo mass $M$ of the source, and parameters that
characterize the production of ionizing photons by the galaxy.  The
first parameter, $dN_{\gamma}/dt$, is the total rate at which hydrogen
ionizing photons from the galaxy enter the IGM. The second is the age
of the source, $t_S$, namely the period of time during which the source
has been active (assuming a steady $dN_{\gamma}/dt$ during that time,
for simplicity).

We consider ages in the range $t_S \sim 10^7$-$10^8$ yr, comparable
to the dynamical time within the host galaxies of interest.  For given 
values of $M$ and $t_S$, we find it useful to express $dN_{\gamma}/dt$
as follows: \be \frac{dN_{\gamma}}{dt} = \frac{M}{m_p}\,
\frac{\Omega_b}{\Omm}\, \frac{\Ni}{t_S}\ , \label{eq-dNdt} \ee where
$m_p$ is the proton mass, and $\Ni$ gives the overall number of
ionizing photons per baryon in the galaxy (ignoring helium). To derive
the expected range of possible values for $\Ni$, we note that it is
determined by \be \Ni = N_{\gamma} \, f_*\, f_{\rm esc}\ , \ee where
we assume that baryons are incorporated into stars with an efficiency
of $f_*$, that $N_{\gamma}$ ionizing photons are produced per baryon
in stars, and that the escape fraction (into the IGM) for the
resulting ionizing radiation is $f_{\rm esc}$. When necessary, we
adopt $f_*=10\%$ (consistent with a comparison of numerical
simulations to the observed cosmic star formation rate at low redshift
[Hernquist \& Springel 2003]), and consider $f_{\rm esc}$ in the full
range 0-1. As explained below, our main conclusions are insensitive to
our choices for the values of $f_*$ and $f_{\rm esc}$.

The parameter $N_{\gamma}$ depends on the initial mass function (IMF)
of the stars.  In what follows, we consider two examples. The first is
the locally-measured IMF of \citet{scalo}, which yields [using
\citet{Leith99}] $N_{\gamma}=4300$ for a metallicity of $1/20$ of the
solar value. The second is an extreme Pop III IMF, assumed to consist
entirely of zero metallicity $M \ga 100\, M_{\odot}$ stars, which
yields $N_{\gamma}=44000$ based on the ionization rate per star, the
stellar spectrum, and the main-sequence lifetime of these stars
\citep{Bromm}. Note that the star formation rate (hereafter SFR)
corresponding to eq.~(\ref{eq-dNdt}) is \be {\rm SFR}=0.171\, \left(
\frac{M}{10^9\, M_{\odot}}\right)\, \left( \frac{t_S} {10^8\, {\rm
yr}} \right)^{-1}\, \left( \frac{f_*} {0.1}\right)\, \left(
\frac{\Omega_b/\Omm}{0.17} \right) \ M_{\odot}\, {\rm yr}^{-1}.
\label{eq-SFR} 
\ee

\section{Ionization and the Damping Wing}

An ionizing source embedded within the neutral IGM can ionize a region
of maximum physical size
\be R_{\rm max} = 0.138\, \left( \frac{M} {10^9\, M_{\odot}} \right)^
{1/3}\, \left( \frac{1+z} {11} \right)^{-1}\, \left( \frac{\Omega_m
h^2} {0.14} \right)^{-1/3}\, \left( \frac{N_{\gamma} f_*} {430}
\right)^{1/3}\, f_{\rm esc}^{1/3}\ {\rm Mpc}\ , \label{eq-Rmax} \ee 
assuming that recombinations are negligible and that all
ionizing photons are absorbed by hydrogen atoms. In this {\it Letter} we
conservatively adopt this maximum size, since a smaller region would
only strengthen our main conclusions (for more elaborate
discussions, see Haiman 2002; Santos 2003).

When dealing with \Lya emission and absorption, it is useful to
associate a redshift $z$ with a photon of observed wavelength
$\lambda_{\rm obs}$ according to $\lambda_{\rm obs}=
\lambda_{\alpha}(1+z)$, where the \Lya wavelength is $\lambda_{\alpha}=
1215.67 $\AA. The optical depth $\tau_{\rm damp}$ owing to the IGM
damping wing of neutral gas between $z_1$ and $z_2$ (where $z_1 < z_2
< z$) can be calculated analytically
\citep{jordi98}. For a source above the reionization redshift, $z_2$
is the redshift corresponding to the blue edge of its \ion{H}{2}
region, and $z_1$ is the reionization redshift. We define $\Delta
\lambda = \lambda_{\alpha}\, (z-z_2)$, which is related to the 
physical distance $\Delta D$ between the source and the \ion{H}{2}
region edge by
\be \Delta D = 0.164\,\left( \frac {\Delta \lambda} {10\,\mbox{\AA}}
\right) \, \left( \frac{1+z}{11}\right)^{-5/2}\, \left(
\frac{\Omega_m h^2} {0.14} \right)^{-1/2}\,{\rm Mpc}\ . 
\ee
In the limit where $z_2$ is very close to $z$ while $z_1$ is much
farther away, the damping optical depth can be accurately approximated as
\be 
\tau_{\rm damp} \approx \frac{1.16\, {\rm Mpc}} {\Delta D}\, 
\left( \frac{\Omega_b / \Omm} {0.17} \right)\, \left\{1 - \frac{\Delta 
\lambda} {2.96 \times 10^3 \mbox{\AA}} \frac{11}{1+z}\, \log \left[ 
\frac{2.68 \times 10^4 \mbox{\AA}} {\Delta \lambda}\, \frac{1+z}{11} 
\right] \right\}\ , \label{eq-damp} 
\ee 
which is independent of $z_1$. 

\section{\Ly-emitting Galaxies}

\label{sec-galxs}

Next we discuss the use of \Ly-emitting galaxies to probe reionization.  To
study the detectability of high-redshift galaxies, we must convert the SFR
of a galaxy in a halo of a given mass [see eq.~(\ref{eq-SFR})] to a flux
level for both the \Lya line and the UV continuum. Assuming that $2/3$ of
the ionizing photons that are reprocessed (i.e., that do not escape from
the galaxy) lead to a hydrogen atom in the $2p$ state and thus result in
the emission of a \Lya photon (Osterbrock 1974), the rest-frame conversion
factor between the SFR and the luminosity of the \Lya line is \be L(\Ly) =
1.8\times 10^{42}\, (1-f_{\rm esc})\, \left( \frac {N_{\gamma}} {4300}
\right)\, \left( \frac{\rm SFR} {1\, M_{\odot}\ {\rm yr}^{-1}} \right)\,
{\rm erg\ s}^{-1}\ , \label{eq-line} \ee where this conversion is
valid for any $t_S \ge 10^7$ yr (since the high-mass, short-lived
stars dominate the ionizing intensity). Using the photon production
rate in \citet{Leith99} for the \citet{scalo} IMF, and the stellar
spectrum and lifetime in Bromm et al.\ (2001) for the extreme Pop III
IMF, we find that the continuum flux can be converted (in the rest
frame) as follows, depending on $t_S$ and source rest-frame wavelength
$\lambda$:
\be F_{\nu}(\lambda) = 10^{27}\, N_{\rm cont}\, \left( \frac{\rm SFR} 
{1\, M_{\odot}\ {\rm yr}^{-1}} \right)\, {\rm erg\ s}^{-1}{\rm\
Hz}^{-1}\ , \label{eq-cont} \ee where \be N_{\rm cont} = \left\{
\begin{array}{ll} 10.\ ,&\mbox{for a \citet{scalo} IMF,}\
t_S=10^8\,{\rm yr},\ \lambda=1500 {\mbox \AA}\ , \\ 8.7\ ,&\mbox{for a
\citet{scalo} IMF,}\ t_S=10^8\,{\rm yr},\ \lambda=2000 {\mbox \AA}\ , \\ 
5.5\ ,&\mbox{for a \citet{scalo} IMF,}\ t_S=10^7\,{\rm yr},\
\lambda=1500 {\mbox \AA}\ , \\ 4.6\ ,&\mbox{for a \citet{scalo} IMF,}\ 
t_S=10^7\,{\rm yr},\ \lambda=2000 {\mbox \AA}\ , \\ 5.9\ ,&\mbox{for
an extreme Pop III IMF,}\ t_S\ge 3 \times 10^6\,{\rm yr},\
\lambda=1500 {\mbox \AA}\ , \\ 3.8\ ,&\mbox{for an extreme Pop III IMF,}\
t_S\ge 3 \times 10^6\,{\rm yr},\ \lambda=2000 {\mbox \AA}\ . \end{array}
\right. \ee Note that for a given SFR and $f_{\rm esc}$, an extreme Pop III
population produces a \Lya line $\sim 10$ times more luminous than does a
\citet{scalo} IMF, but the UV continuum is comparable for the two
IMFs.  The reason is that high-mass, short-lived stars strongly dominate
the production of ionizing photons, but in the Scalo IMF case, lower-mass
and longer-lived stars contribute substantially to the continuum.

To evaluate the impact of absorption on the \Lya line, we require an
estimate of the width of the line. Assuming a Gaussian profile, we
estimate the one-dimensional velocity dispersion $\sigma_v$ by
$\sigma_v \approx V_c/\sqrt{2}$, where $V_c$ is the virial circular
velocity of the halo; this yields
\be \sigma_v = 26.9\, \left(\frac{M}{10^9\, M_{\odot}} \right)^{1/3}\, 
\left(\frac{1+z}{11} \right)^{1/2}\, \left(\frac{\Omm h^2}{0.14} 
\right)^{1/6}\, {\rm km\ s}^{-1}\ . \label{eq:sigv} \ee The observed 
scale-length $\sigma_{\lambda}$ is related to $\sigma_v$ by
\be \sigma_v = 22.4\, \left( \frac{\sigma_{\lambda}} {1\,\mbox{\AA}} 
\right)\, \left(\frac{1+z}{11} \right)^{-1}\, {\rm km\ s}^{-1}\ . \ee

\section{Results}

Identifying the redshifted Ly$\alpha$ line based on its properties and a
strong photometric continuum break, \citet{z10} measured an amplified line
flux for the candidate $z=10.0$ galaxy of $(4.1\pm 0.5) \times 10^{-18}$
erg cm$^{-2}$ s$^{-1}$ and a UV continuum flux density of
$(3.6^{+1.0}_{-0.7}) \times 10^{-30}$ erg cm$^{-2}$ s$^{-1}$ Hz$^{-1}$
averaged over the 1360-1640\AA\ rest-frame band, and $(2.3^{+0.9}_{-0.7})
\times 10^{-30}$ erg cm$^{-2}$ s$^{-1}$ Hz$^{-1}$ averaged over the
1830-2090\AA\ rest-frame band. Correcting for the cluster lensing
amplification factor of $\sim 25$, they estimated a SFR of 2-3 $M_{\odot}$
yr$^{-1}$ based on the continuum strength and 0.03-0.09 $M_{\odot}$
yr$^{-1}$ based on the line flux. This discrepancy, they suggested,
indicates a factor of $\sim 40$ absorption of the line, in principle owing
to both \ion{H}{1} absorption and dust extinction, but in this case they
noted that the steep UV slope indicates negligible extinction. The observed
line is somewhat resolved, allowing \citet{z10} to place an upper limit of
$\sim 200$ km/s on $\sigma_v$, with best fits obtained for $\sigma_v < 60$
km/s.

We reconsider the interpretation of the observations of \citet{z10},
accounting for different possible IMFs and source ages. We find a
luminosity distance to the galaxy of 105 Gpc, yielding a rest-frame
luminosity (demagnified by a factor of 25) of $2.2 \times 10^{41}$ erg
s$^{-1}$ in the line and a (demagnified) rest-frame luminosity density
of $1.7 \times 10^{28}$ erg s$^{-1}$ Hz$^{-1}$ and $1.1 \times
10^{28}$ erg s$^{-1}$ Hz$^{-1}$ in the two UV continuum ranges,
respectively (including the extra redshift factor for frequency
bands). Estimating the overall conversion factor with the values for
$\lambda=1500$\AA\ and for $\lambda=2000$\AA,\, respectively, we find
using eq.~(\ref{eq-cont}) the SFR implied by the continuum flux, which
then implies through eq.~(\ref{eq-line}) an intrinsic \Lya line
luminosity (before absorption). We also estimate a halo mass using
eq.~(\ref{eq-SFR}), assuming $f_* = 0.1$.

Table~1 shows our derived parameters for the $z=10.0$ galaxy and its
host halo, depending on the IMF and the source lifetime. The Table
gives the intrinsic line luminosity expected in the absence of
\ion{H}{1} absorption or dust extinction; also given are the expected
intrinsic $\sigma_v$ based on eq.~(\ref{eq:sigv}), and the maximum
physical size of the \ion{H}{2} region based on
eq.~(\ref{eq-Rmax}). Given the observations, the derived quantities
given in the Table are independent of the star formation efficiency
$f_*$, except for the estimated halo mass and $\sigma_v$. Note that
the observed ratio of the values of the continuum flux in the two
measured wavelength regions is $1.6 \pm 0.7$, compared to a predicted
ratio of 1.2 for a Scalo IMF and 1.6 for a Pop III IMF. Thus, the
steep UV slope is more consistent with a Pop III IMF, but the
discrepancy with a Scalo IMF is not highly significant. In either
case, this measurement suggests that dust extinction is not strong
\citep[compare discussion in][]{z10}. For each case in the Table we 
average the SFR derived from the two UV continuum measurements. The
errors in these two measurements introduce errors of $\sim 20\%$ in
the predicted \Lya line luminosity and $\sim 7\%$ in the predicted
$R_{\rm max}$.

\begin{deluxetable}{llclcccc}
\label{tab-z6p6}
\tablenum{1}
\footnotesize
\tablecaption{Derived parameters for the \citet{z10} galaxy (see text)}
\tablecolumns{8}
\tablehead{
\colhead{IMF} &
\colhead{$t_{S}$} &
\colhead{SFR} &
\colhead{Halo mass} &
\colhead{\Lya luminosity} &
\colhead{$\sigma_v$} &
\colhead{$R_{\rm max}$} &
\colhead{$L^{\alpha}_{\rm max}$} \\
\colhead{} &
\colhead{[yr]} &
\colhead{[$M_{\odot}$ yr$^{-1}$]} &
\colhead{[$M_{\odot}$]} &
\colhead{[$(1-f_{\rm esc})$ erg s$^{-1}$]} &
\colhead{[km/s]} &
\colhead{[$f_{\rm esc}^{1/3}$ Mpc]} &
\colhead{[erg s$^{-1}$]}
}
\startdata
Scalo & $10^7$ & 2.8 & $1.6\times 10^9$ & $5.0 \times 10^{42}$ 
& 32 & 0.16 & $5.1 \times 10^{38}$ \\
Scalo & $10^8$ & 1.5 & $8.8\times 10^9$ & $2.7 \times 10^{42}$ 
& 56 & 0.28 & $9.4 \times 10^{39}$ \\
Pop III & $10^7$ & 2.9 & $1.7\times 10^9$ & $5.3 \times 10^{43}$ 
& 32 & 0.36 & $5.5 \times 10^{41}$ \\
Pop III & $10^8$ & 2.9 & $1.7\times 10^{10}$ & $5.3 \times 10^{43}$ 
& 69 & 0.77 & $4.8 \times 10^{42}$ \\
\enddata
\end{deluxetable}

For a given value of $f_{\rm esc}$, the rest-frame \Lya line
luminosity can be predicted from the UV continuum by accounting for
the factor of $1-f_{\rm esc}$ (which arises since ionizing photons
produce \Lya only if they are absorbed within the galaxy and
reprocessed), and adding the damping wing absorption which depends on
$R_{\rm max}$ (which depends on $f_{\rm esc}$ as well). For each case
in the Table, we vary $f_{\rm esc}$ over all possible values and find
the maximum possible \Lya line luminosity $L^{\alpha}_{\rm max}$. If
we allow a 2-$\sigma$ measurement error in the UV continuum flux, then
the value of $L^{\alpha}_{\rm max}$ becomes, respectively for the four
cases shown in Table~1, $2.1 \times 10^{39}$ erg s$^{-1}$, $2.6 \times
10^{40}$ erg s$^{-1}$, $1.4 \times 10^{42}$ erg s$^{-1}$, and $9.3
\times 10^{42}$ erg s$^{-1}$. These numbers can be compared to the
observed value of $2.2 \times 10^{41}$ erg s$^{-1}$ (with a 2-$\sigma$
lower limit of $1.7 \times 10^{41}$ erg s$^{-1}$). Thus, the
\citet{scalo} IMF predicts a much weaker line than that observed and is
ruled out, assuming that the IGM surrounding the \ion{H}{2} region is
neutral. 
An extreme Pop III IMF, however, is easily consistent with the observations
even for a short burst ($t_S=10^7$ yr).

Each line in the Table provides the mass of the source
halo. Interestingly, numerical simulations indicate that the
multiplicity function of star formation at $z\sim 10$ peaks at halo
masses $\sim 10^9 M_{\odot}$ (Springel \& Hernquist 2003), comparable
to our inferred values. Using a halo mass function based on numerical
simulations \citep{shetht99, shetht02}, we find that the mean physical
separation between halos at or above the inferred mass is 0.21 Mpc,
0.54 Mpc, 0.21 Mpc, and 0.82 Mpc, respectively, for the four cases in
the Table.

We have found that the \citet{scalo} IMF is inconsistent with the
observations if the \ion{H}{2} region is produced by the source galaxy
and is surrounded by fully neutral IGM. There are two possible ways to
resolve this inconsistency. One is to assume a larger surrounding
\ion{H}{2} region that is produced by the collective ionizing flux
from nearby galaxies. The required minimum size
for producing the observed line flux (with $f_{\rm esc}=0$ in order to
maximize the intrinsic line flux) is 0.35 Mpc ($t_S=10^7$ yr) or 0.43
Mpc ($t_S=10^8$ yr). Given the above halo abundances, such a sphere
will typically contain a number of equally or more massive halos,
i.e., an average number of 20 and 2, respectively. \citet{Nbody}
showed that the fluctuations in the number of halos in such a volume
are much larger than Poisson fluctuations, owing to biased galaxy
formation which greatly amplifies large-scale density perturbations.
In particular, the typical 1-$\sigma$ fluctuation in the number of
halos is a factor of 3 and 5 in these two cases, respectively; thus it
is possible that the selection effect of the observability of the \Lya
line has revealed a galaxy in a region that had been previously
reionized, e.g., by a positive 2-$\sigma$ fluctuation in the local
number density of galaxies, while most of the rest of the universe
remained neutral at that redshift.

\begin{figure}[htbp] 
\plotone{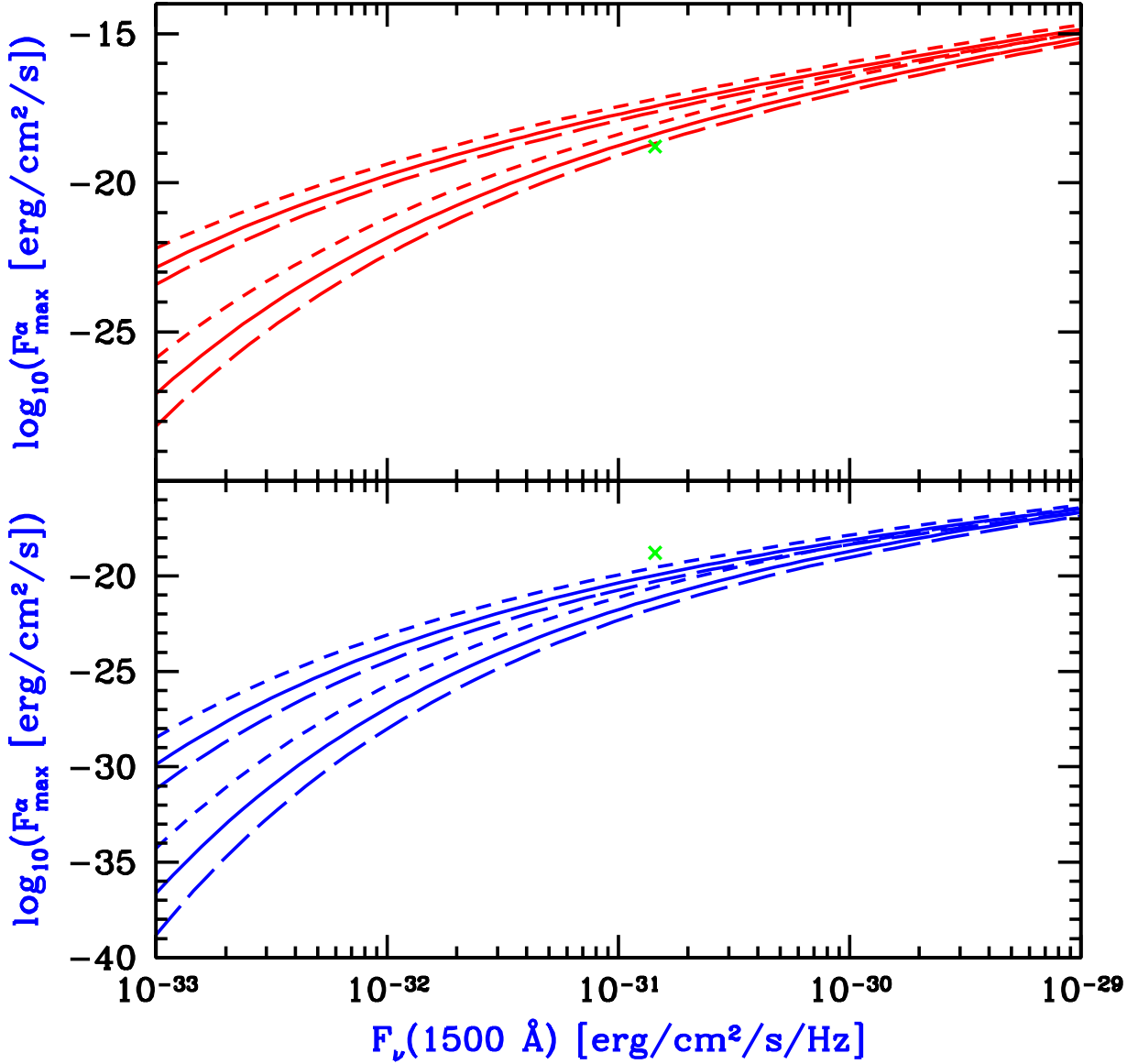} 
\caption{Maximum observable flux of the \Lya line versus UV continuum flux
(at a rest-frame wavelength of 1500\AA) of a galaxy prior to
reionization (i.e., assuming that the \ion{H}{2} region of the galaxy
is surrounded by a neutral IGM). We assume a narrow emission line
($\la 100~{\rm km~s^{-1}}$). We consider a \citet{scalo} IMF (bottom
panel) or a Pop III IMF (top panel), at redshifts 7 (short-dashed
curves), 10 (solid curves), or 13 (long-dashed curves). In each case,
a pair of curves are shown, corresponding to $t_S = 10^8$ yr (higher)
or $t_S = 10^7$ yr (lower). Also shown are the observed flux values
for the
\citet{z10} $z=10$ galaxy, demagnified by a factor of 25 (denoted by
{\bf $\times$}).}
\label{fig} 
\end{figure}

Another possibility is that the \ion{H}{2} region has the size given in the
Table, but the IGM beyond the \ion{H}{2} region is on average only
partially ionized and thus produces a weaker damping wing than we have
assumed thus far. If as before we maximize the flux over all possible
values of $f_{\rm esc}$, then the maximum allowed neutral fraction is 0.27
($t_S=10^7$ yr) or 0.37 ($t_S=10^8$ yr).

The damping wing of a neutral IGM limits the maximal flux of the \Lya
line from any galaxy prior to reionization.  Figure 1 illustrates this
upper limit as a function of the UV continuum flux of that galaxy,
which affects the size of the \ion{H}{2} region around it. We have
maximized the line flux over $f_{\rm esc}$ and calculated the damping
wing at the center of the line assuming that the IGM is neutral and
that its opacity does not change significantly across the width of the
line. The latter assumption is well satisfied for lines with velocity
widths $\la 100~{\rm km~s^{-1}}$. Our flux limits are independent of
the star formation efficiency $f_*$. The galaxy discovered by
\citet{z10} (denoted by {\bf $\times$}) should be compared to the
solid lines (indicating results for $z=10$);
its detected \Lya line flux requires that the IGM be ionized for a
\citet{scalo} IMF (bottom panel) but is consistent with a neutral IGM for a
Pop III IMF (top panel).

\section{Discussion}

Figure 1 shows that if the galaxy discovered by \citet{z10} indeed has
a redshift $z=10$, then either its stars are very massive ($\ga
100M_\odot$) and hence capable of creating a large \ion{H}{2} region
around it, or the large-scale IGM around it has already been mostly
reionized (with a neutral fraction $\la 0.4$). The ratio of the
galaxy's UV continuum flux in two bands is consistent with the first
possibility, but better determination of the continuum spectrum or
detection of other recombination lines (Bromm, Kudritzki, \& Loeb
2001; Tumlinson, Giroux, \& Shull 2001; Oh, Haiman, \& Rees 2001) is
required to make a convincing case.  For example, a stellar population
with a top-heavy IMF will produce substantial flux in He II Ly$\alpha$
($303.9 $\AA) and He II $\lambda 1640$.  A search for features in the
spectrum of a redshift $z=10$ galaxy at $\lambda_{\rm obs}= 3343$\AA\
and $1.8\mu$m would constrain the first interpretation.
The discovery of other galaxies may test the second possibility, since the
damping wing of a neutral IGM limits the maximum \Lya line flux of any
galaxy prior to reionization based on its UV continuum flux.

The imprint of the damping wing is insensitive to a variety of complicating
factors. Clumping of gas and its infall onto the galaxy have a minor
effect, since the damping wing averages the IGM over large physical scales
($\sim 1$ Mpc).  Moreover, the characteristic size of the \ion{H}{2} region
around a bright galaxy at $z\sim 10$ is an order of magnitude larger than
the scale of its infall region. The characteristic peculiar velocity of
galaxies at $z\sim 10$ is an order of magnitude smaller than the Hubble
velocity at the edge of their \ion{H}{2} regions, and can only
significantly change the effect of resonant absorption on the line
profile. We have ignored resonant absorption from \ion{H}{1} within the
\ion{H}{2} region since it would not suppress even the blue wing of the
\Lya emission line if the galaxy had a peculiar velocity (in the direction
of recession) larger than the line width (thus allowing the line to appear
nearly symmetric); such a peculiar velocity may not be improbable within
small samples because it makes the line more easily observable and adds
a selection bias.  In any case, the addition of resonant absorption would
only strengthen our limits on the \Lya line flux before reionization. Dust
extinction in the host galaxy would affect ionizing photons (and thus the
\Lya line flux) more than it would the continuum at longer
wavelengths. Based on the observed UV continuum, we would have predicted a
weaker line with the inclusion of extinction, and so our argument would
have been stronger.

We have considered the lowest (most probable) value within the range of
lensing magnification factors inferred by \citet{z10}. A higher value would
again strengthen our limits.  
For example, a magnification factor of 100 would imply that the line
was 4 times weaker intrinsically, but the predicted line from the UV
continuum would then be weakened by more than a factor of 4 since the
smaller \ion{H}{2} region would imply a damping wing that suppressed
more of the line flux.

Current observational constraints on reionization provide an
inconsistent picture.  On the one hand, the large-scale polarization
anisotropies of the cosmic microwave background measured by WMAP imply
a reionization redshift of 10-20 (Kogut et al.\ 2003), while on the
other hand, the extent of the \ion{H}{2} regions around the highest
redshift quasars indicates\footnote{The evidence for a possible
Gunn-Peterson (1965) trough in these quasar spectra (White et
al. 2003) is still being debated (Songaila 2004).} a significantly
neutral IGM at $z\sim 6.4$ (Wyithe \& Loeb 2004). Moreover, the IGM
should have been cooler than observed at $z\sim 3$-4 if hydrogen had
been fully reionized at $z\ga 9$ (Theuns et al.\ 2002; Hui \& Haiman
2003). Theoretically, it is possible that the ionization fraction
evolved in a complex, non-monotonic fashion owing to an early episode
of Pop III star formation (Wyithe \& Loeb 2003; Cen 2003).

Existing observations do not enable us to discriminate between the two
interpretations of a positive detection of a \Lya emission line from the
$z=10$ galaxy of \citet{z10}. Sokasian et al.\ (2003a,b) have studied the
process of reionization using cosmological simulations with radiative
transfer; they have shown that for a Scalo-type IMF, reionization occurs
over an extended redshift interval.  For models in which the \Lya optical
depth at $z\approx 6$ matches the value inferred from the SDSS quasars, a
substantial fraction of the mass in the universe ($>30 \%$) is already
completely ionized by $z=10$.  Since reionization proceeds ``inside-out''
affecting overdense regions first within the simulated volume (see also
Gnedin 2000), we would expect the environment of the \citet{z10} galaxy to
be ionized.  If the evolution had been more complex and included sources
with population III stars prior to $z=12$, an even larger fraction of the
mass would have been ionized by $z=10$ (Wyithe \& Loeb 2003; Sokasian et
al.\ 2003b). This picture could change, however, if reionization occurred
``outside-in'' (e.g., Miralda-Escud\'e, Haehnelt \& Rees 2000). As shown by
Barkana \& Loeb (2004), the fluctuations in the real universe should be far
larger than indicated in existing simulations because of their limited box size.
In reality, small fluctuations on scales larger than the box are greatly
amplified through biased galaxy formation at high redshifts.

The various scenarios for reionization can be tested with more data on
\Lya emitting galaxies (supplementing recent work by Hu et al. 2002,
2004; Malhotra \& Rhoads 2002; Kodaira et al.\ 2003; Santos et al.\
2003; Kneib et al.\ 2004; Stanway et al.\ 2004; Barton et al.\ 2004),
SDSS quasars (Fan et al.\ 2003), or gamma-ray burst afterglows (Loeb
2003; Barkana \& Loeb 2004a).
Over the next decade, it may also be possible to map directly the
neutral hydrogen in the IGM through its $21\,$cm line (see recent
discussions by Zaldarriaga, Furlanetto, \& Hernquist 2003; Morales \&
Hewitt 2003; Loeb \& Zaldarriaga 2003; Gnedin \& Shaver 2003;
Sokasian, Furlanetto \& Hernquist 2003) with forthcoming instruments
such as LOFAR ({\it http://www.lofar.org/}) or SKA ({\it
http://www.skatelescope.org/}).

\acknowledgments

A.L. and R.B. acknowledge support by NSF grant AST-0204514 and NATO
grant PST.CLG.979414. R.B. is grateful for the support of an Alon
Fellowship at Tel Aviv University and of Israel Science Foundation
grant 28/02/01. This work was also supported in part by NSF grant
AST-0071019 and NASA grant NAG 5-13292 (for A.L.).

\end{document}